# On Enforcing Satisfiable, Coherent, and Minimal Sets of Self-Map Constraints in *MatBase*

**Christian Mancas***

*Mathematics & Computer Science Department, Ovidius University at Constanta, Romania*

***Corresponding Author:** Christian Mancas, Mathematics & Computer Science Department, Ovidius University, Bd. Pipera 1/U, Voluntari, IF, Romania.

## Abstract

This paper rigorously and concisely defines, in the context of our (Elementary) Mathematical Data Model ((E)MDM), the mathematical concepts of self-map, composite mapping, totality, one-to-oneness, non-primeness, ontoness, bijectivity, default value, (null-)reflexivity, irreflexivity, (null-)symmetry, asymmetry, (null-)idempotency, anti-idempotency, (null-)equivalence, acyclicity, (null-)representative system mapping, the properties that relate them, and the corresponding corollaries on the coherence and minimality of sets made of such mapping properties viewed as database constraints. Its main contribution is the pseudocode algorithm used by *MatBase*, our intelligent database management system prototype based on both (E)MDM, the relational, and the entity-relationship data models, for enforcing self-map, atomic, and composite mapping constraint sets. We prove that this algorithm guarantees the satisfiability, coherence, and minimality of such sets, while being very fast, solid, complete, and minimal. In the sequel, we also presented the relevant *MatBase* user interface as well as the tables of its metacatalog used by this algorithm.

## Abbreviations

DBMS = Database Management System.
db(s) = database(s).
(E)MDM = (Elementary) Mathematical Data Model.
E-R = Entity-Relationship.
iff = if and only if.

## Introduction

We presented in [1] the current version of our (Elementary) Mathematical Data Model ((E)MDM). Out of its 76 constraint types, there are 6 pertaining to all mappings, namely totality, one-to-oneness, non-primeness, ontoness, bijectivity, and default value, and 14 pertaining only to self-maps, which are particular cases of dyadic relations [2]: (null-)reflexivity, irreflexivity, (null-)symmetry, asymmetry, (null-)idempotency, anti-idempotency, (null-)equivalence, (null-)representative system mapping, and acyclicity. As usual in mathematics, some of them or some combinations of them imply others, while some of them are mutually exclusive. This is why any intelligent Database Management System (DBMS) must accept only satisfiable, coherent, and, for optimality concerns, also minimal sets of constraints.

*MatBase* [3] is our intelligent DBMS prototype, based on both (E)MDM, the Entity-Relationship (E-R) Data Model [4, 5, 6], the Relational Data Model [6, 7, 8], and Datalog¬ [8, 9], currently implemented in two MS platforms: Access (for small dbs and undergraduate students) and .NET C# and SQL Server (for large dbs and MSc. students). Its (E)MDM interface provides users with a form (see, e.g., Figure 1) in which all metadata [10] for any mapping of a database (db) it manages may be inspected and updated. Please note that, for composite mappings (e.g., the self-map *State* ° *StateCapital* from Figure 1), the *FUNCTIONS* sub-form (which manages all mappings defined on the current set) has a subform that manages the corresponding member mappings (e.g., *State* : *CITIES* → *STATES* and *StateCapital* : *STATES* ↔ *CITIES* from Figure 1).

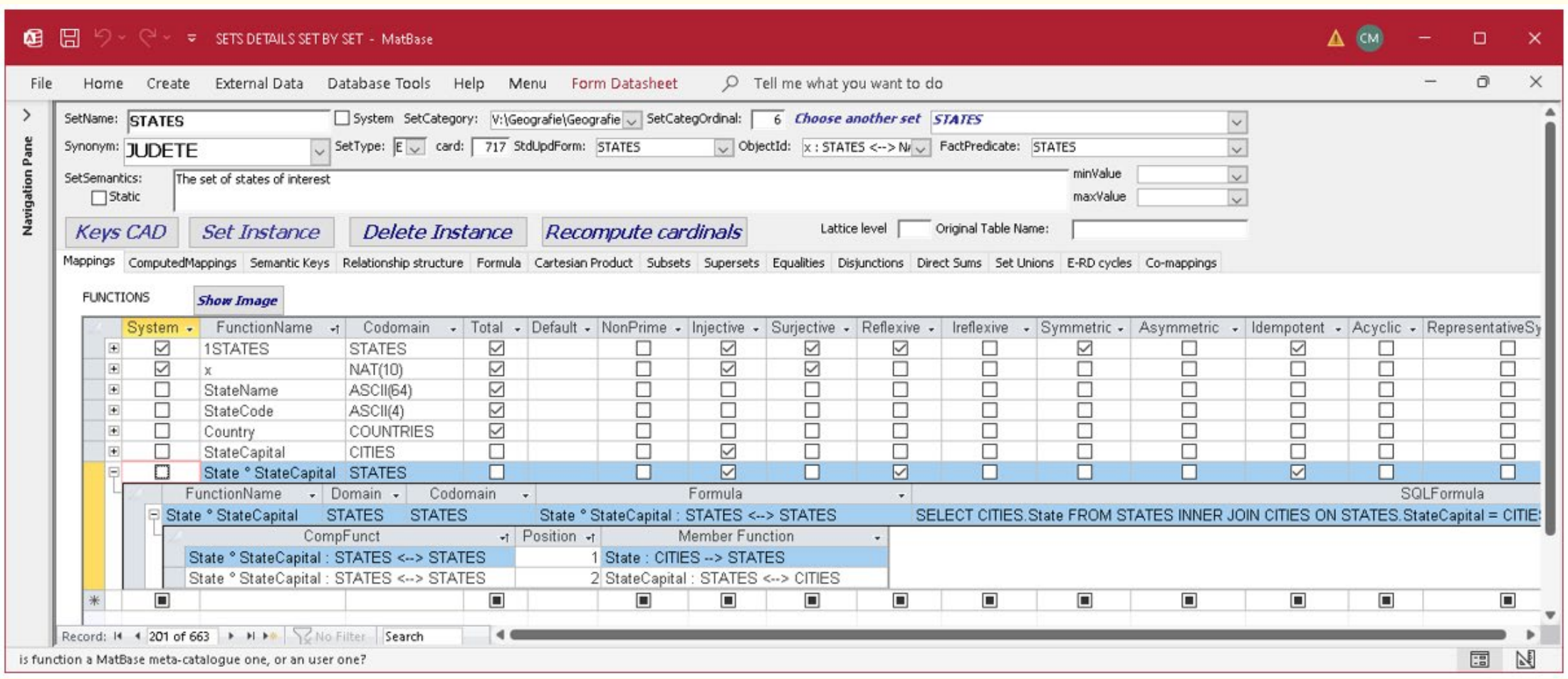

***Figure 1:*** MS Access *MatBase* form for managing db sets' schema.

In particular, for self-maps users may assert or delete their properties by simply clicking on the corresponding checkboxes from the *Functions* tab. Immediately after each such click, *MatBase* analyzes the new desired such constraint set and undoes the update if it is invalid (e.g., the current function is not a self-map one, or the corresponding constraint set would be incoherent, or the user tried to delete a redundant constraint, or the current db instance does not satisfy the newly desired constraint set, etc.). If the update is valid, then *MatBase* not only accepts it, but also automatically updates the subset of corresponding redundant constraints and generates or deletes the code needed to enforce the newly desired mapping type constraint set.

This paper describes the math behind this process, as well as the metadata and algorithm that *MatBase* uses to perform these tasks. Of course, that 17 of these 20 constraint types are non-relational, i.e., they may not be enforced by any relational DBMS (e.g., MS SQL Server, Oracle Database, IBM DB2, etc.): the only relational ones are totality (NOT NULL), one-to-one-ness (UNIQUE), and default values (DEFAULT). Consequently, the 17 non-relational ones should be enforced by db software applications managing the corresponding relational dbs. *MatBase* automatically generates such software applications for every db it manages.

## Related work

*MatBase's* constraint sets coherence and minimality enforcement algorithms were generally presented at a higher conceptual level in [11]. First, [11] deals with all (E)MDM constraint types (which were only 61 at that time); then, it does not address the particularities of self-maps, which are cases of dyadic relations (which are cases of homogeneous binary function products (i.e., of type $f \bullet g : D \to (C \cup NULLS)^2$, where NULLS is a distinguished countable set of *null values*), for which the first canonical Cartesian projection is the unity function of the corresponding domain (i.e., of type $\mathbf{1}_D: D \to D$, $\mathbf{1}_D(x) = x$, $\forall x \in D$) and the second one is the functional dyadic relation, which might not be totally defined (i.e., it may take null values as well, $f: D \to (D \cup \text{NULLS})$). Moreover, [11] does not deal either with rejecting sets of constraint types that would duplicate the unity mappings of the corresponding object sets.

Deeper details on self-maps (autofunctions) enforcement in *MatBase* were presented in [12, 13].

Proofs of the mathematical results presented in the next section may be found, e.g., in [9, 14].

(E)MDM is also a 5th generation programming language [15, 16] and *MatBase* is also a tool for transparent programming while modeling data at conceptual levels [3].

To our knowledge, the other most closely related approaches to non-relational constraint enforcement are based on business rules management (BRM) [17, 18] and their corresponding implemented systems (BRMS) and process managers (BPM), like the IBM Operational Decision Manager [19], IBM Business Process Manager [20], Red Hat Decision Manager [21], Agiloft Custom Workflow/BPM [22], etc. They are generally based on XML (but also on the Z notation, Business Process Execution Language, Business Process Modeling Notation, Decision Model and Notation, or the Semantics of Business Vocabulary and Business Rules), which is the only other field of endeavor trying to systematically deal with business rules, even if informally, not at the db design level but at the software application one, and without providing automatic code generation.

From this perspective, (E)MDM is also a BRM but a formal one, and *MatBase* is also a BRMS but an automatically code generating one.

The satisfiability, coherence, and minimality of first order predicate formulae sets has been extensively studied mathematically (e.g., [23]) but not in the db contexts, as there are only six relational constraint types (for which any combination is coherent), out of which NoSQL DBMSes only use 2 or 3.

## Materials and Methods

The following definitions, propositions, and corollaries are from Appendix A ("The Math Behind (E)MDM") of [9]. The propositions are from its subsections A.3.2.2 ("Self-maps") and A.3.2.3 ("Partially defined self-maps"), while the corollaries are from its section A.6 ("Coherence and Minimality of Mapping Constraint Sets").

### *Definitions*

0. a. *A relation R* is a subset of a Cartesian product of *n* sets (not necessarily distinct), $n > 1$, natural: $R \subseteq S_1 \times \ldots \times S_n$.
   b. *A dyadic relation R* is a subset of a binary Cartesian product of a set *S* with itself: $R \subseteq S \times S$.
   c. A dyadic relation R is *left unique* iff $\forall x_1, x_2, y \in S$, $x_1Ry \wedge x_2Ry \Rightarrow x_1 = x_2$.
   d. A dyadic relation R is *right unique* (*functional*) iff $\forall x, y_1, y_2 \in S$, $xRy_1 \wedge xRy_2 \Rightarrow y_1 = y_2$.
   e. A dyadic relation R is *left serial* iff $\forall y \in S, \exists x \in S, xRy$.
   f. A dyadic relation R is *right serial* iff $\forall x \in S, \exists y \in S, xRy$.
1. a. A right unique and right serial dyadic relation *sm* over *S* is called a *self-map* (*autofunction*) and is denoted $sm : S \to S$, with $sm(x) = y$, instead of $x\ sm\ y$, $\forall x, y \in S$. For self-maps, right seriality is called *totality* and right uniqueness is called *functionality*.
   b. A self-map that is not right serial is called *partial*(*ly defined*) and is denoted $sm : S \to S \cup NULLS$, where NULLS is a distinguished countable set of *null-values*.

c. A left and right unique dyadic relation *sm* over *S* is called a *one-to-one* (*injective*) *self-map* (*autofunction*) and is denoted $sm : S \leftrightarrow S$.

d. A right unique and left serial dyadic relation *sm* over *S* is called a *onto* (*surjective*) *self-map*.

e. Any binary functional relation $f \subseteq D \times C$ (over sets *C* and *D*) is called a *mapping* (*function*) and is denoted $f : D \rightarrow C$ (where *D* is called its *domain*, and *C* its *codomain*); the *image of f* is the set $Im(f) = \{y \mid \exists x \in D, f(x) = y\} \subseteq C$; for any proper subset $B \subset D$, $f|_B : B \rightarrow C$ is called the *restriction of f* to *B*; trivially, self-maps are mappings with $D = C$ or $D \subseteq C$ or $C \subseteq D$.

f. A mapping $f : D \rightarrow C$ is *total* (*totally defined*) iff $C \cap$ NULLS = ∅; otherwise, *f* is *partially defined*.

g. For any set *S* there is a unique distinguished associated self-map called its *unity mapping* and denoted $\mathbf{1}_S : S \leftrightarrow S$, defined as $\mathbf{1}_S(x) = x, \forall x \in S$. If $S \subseteq T$, $\mathbf{1}_S = \mathbf{1}_T|_S$ is also called the associated *canonical injection* mapping.

h. A mapping $f : D \rightarrow C$ is *one-to-one* (*injective*) iff $f(x_1) = y = f(x_2) \Rightarrow x_1 = x_2, \forall x_1, x_2 \in D$.

i. A (*Cartesian*) *mapping product* $f_1 \bullet \ldots \bullet f_n : D \rightarrow C_1 \times \ldots \times C_n$, $n > 1$, natural (called *arity*), is the mapping defined by $(f_1 \bullet \ldots \bullet f_n)(x) = (y_1, \ldots, y_n), \forall x \in D, \forall y_i \in C_i, 1 \leq i \leq n > 1$, naturals. If $n = 1$, $f_1 : D \rightarrow C_1$ is called *atomic*.

j. A (Cartesian) mapping product $f_1 \bullet \ldots \bullet f_n : D \rightarrow C_1 \times \ldots \times C_n$, $n > 1$, natural, is *minimally one-to-one* iff it is one-to-one (i.e., $(f_1 \bullet \ldots \bullet f_n)(x_1) = (y_1, \ldots, y_n) = (f_1 \bullet \ldots \bullet f_n)(x_2) \Rightarrow x_1 = x_2, \forall x_1, x_2 \in D, \forall y_i \in C_i, 1 \leq i \leq n > 1$, naturals) and none of its proper subproducts is one-to-one.

k. A mapping $f : D \rightarrow C$ is *non-prime* if it is neither one-to-one, nor a member of a minimally one-to-one (Cartesian) mapping product.

l. A relation $R \subseteq S_1 \times \ldots \times S_n$, $n > 1$, natural, may also be viewed as the one-to-one (Cartesian) mapping product $f_1 \bullet \ldots \bullet f_n : R \rightarrow S_1 \times \ldots \times S_n$ of its canonical Cartesian projections $f_i : R \rightarrow S_i$, $1 \leq i \leq n$, defined as $f_i(x) = y_i, \forall x \in R, y_i \in S_i$, all of them being totally defined.

m. A mapping $f : D \rightarrow C$ is *onto* (*surjective*) iff $\forall y \in C, \exists x \in D$ such that $f(x) = y$ (i.e., $Im(f) = C$).

n. A mapping *f* is *bijective* iff it is both one-to-one (injective) and onto (surjective).

o. A mapping $f : D \rightarrow C \cup$ NULLS has a *default value* $v \in C$ (denoted *f default v*) iff $\forall x \in D, f(x) \in$ NULLS $\Rightarrow f(x)$ is automatically set to *v* by the DBMS managing updates of *f*.

p. Given mappings $f_n : D \rightarrow S_n, f_{n-1} : S_n \rightarrow S_{n-1}, \ldots, f_2 : S_3 \rightarrow S_2, f_1 : S_2 \rightarrow C$, $n > 1$, natural, the *composite mapping* $cm = f_1 \circ \ldots \circ f_n : D \rightarrow C$ is defined as $cm(x) = (f_1 \circ \ldots \circ f_n)(x) = f_1(f_2(\ldots f_{n-1}(f_n(x))\ldots)), \forall x \in D$ (as it is easy to prove that mapping composition is associative, no parenthesis were used to define it). When $n = 1$, *cm* is called *single* (*not composite*).

r. Given an equivalence dyadic relation ~ over a set *S*, the set *S*/~ of the corresponding *equivalence classes* (*blocks*, *partitions*) is called the *quotient set of S with respect to ~*.

s. Between any set *S* and its quotient set with respect to an equivalence relation ~ there is a unique *canonical surjection* (*onto mapping*) $\rho_\sim : S \rightarrow S/_\sim$, defined as $\rho_\sim(x) = y$, where y is the equivalence class to which *x* belongs, $\forall x \in S$.

2. A self-map *sm* over a set *S* (having any distinct elements *x*, *y*, *z*) is:
   a. *reflexive* iff $sm(x) = x$
   b. *null-reflexive* iff $sm(x) = x \vee sm(x) \in$ NULLS
   c. *irreflexive* iff $sm(x) \neq x$
   d. *symmetric* iff $sm(x) = y \Rightarrow sm(y) = x$
   e. *null-symmetric* iff $sm(x) = y \Rightarrow sm(y) = x \vee sm(y) \in$ NULLS
   f. *asymmetric* iff $sm(x) = y \Rightarrow sm(y) \neq x$
   g. *idempotent* iff $sm(x) = sm(sm(x)) = sm^2(x)$
   h. *null-idempotent* iff $sm^2(x) = sm(x) \vee sm^2(x) \in$ NULLS
   i. *anti-idempotent* iff $sm^2(x) \neq sm(x)$
   j. *equivalence* iff it is both reflexive, symmetric, and idempotent
   k. *null-equivalence* iff it is both (null-)reflexive, (null-)symmetric, and (null-)idempotent
   l. *representative system mapping* (of *S* with *respect to an equivalence relation ~ over it*) iff $sm : S \rightarrow S$, $sm = rs_\sim \circ \rho_\sim$, where $\rho_\sim : S \rightarrow S/_\sim$ is the canonical surjection of *S* with respect to ~ and $rs_\sim : S/_\sim \rightarrow S$ is a mapping that associates to any equivalence class $c \in S/_\sim$ one of its elements $y \in S$ (called the *representative* of that class), i.e., $sm(x) = rs_\sim(\rho_\sim(x)) = rs_\sim(c) = y, \forall x, y \in c \subseteq S$.

m. *null-representative system mapping* (*of S with respect to an equivalence relation ~ over it*) iff $sm : S \to S \cup$ NULLS, where, as for l. above, $sm = rs_{\sim} \circ \rho_{\sim}$, $\rho : S \to S/_{\sim}$, $rs_{\sim} : S/_{\sim} \to S \cup$ NULLS, $sm(x) = rs_{\sim}(\rho_{\sim}(x)) = rs_{\sim}(c) = y \vee sm(x) = rs(c) \in$NULLS, $\forall x, y \in c \subseteq S$.

n. *acyclic* iff $x_2 = sm(x_1) \wedge x_3 = sm(x_2) \wedge \ldots \wedge x_n = sm(x_{n-1}) \Rightarrow x_1 \neq sm(x_n)$, for any natural $n > 0$ and distinct $x_1, \ldots, x_n \in S$.

o. *left-inEuclidean iff* $sm(y) = x = sm(z) \Rightarrow y \neq x \neq z$

p. *Euclidean and inEuclidean* iff $((y = sm(x) \Rightarrow z \neq msm(x)) \vee (z = sm(x) \Rightarrow y \neq sm(x))) \wedge ((x = sm(y) \Rightarrow x \neq sm(z)) \vee (x = sm(z) \Rightarrow x \neq sm(y))) \Leftrightarrow (z = sm(x) \Rightarrow y \neq sm(x)) \wedge (x = sm(y) \Rightarrow x \neq sm(z))$ (as functionality guarantees that both $(y = sm(x) \Rightarrow z \neq sm(x))$ and $(x = sm(z) \Rightarrow x \neq sm(y))) \Leftrightarrow y \neq z \wedge sm(y) \neq sm(z)$

3. A *constraint* is a first order logic formula that has all its variable occurrences bound to a universal quantifier (i.e., ∀–for any– and ∃– there is). For example, all above 20 properties are constraint types of self-maps.
4. A constraint is *satisfied* by a set of values for its variables if it has value *true* for them; otherwise, it is *violated*. A constraint set is *satisfied* by a set of values for all its variables if all its constraints are satisfied.
5. A constraint set is *incoherent* iff it is satisfied only by the corresponding empty set. For example, according to the first order logic laws of *non-contradiction* ("nothing can be both true and false simultaneously") and *excluded middle* ("everything is either true or false, but not neither"), the sets {*sm* reflexive, *sm* irreflexive} and {*sm* symmetric, *sm* asymmetric} are incoherent, for any self-map *sm*.
6. A constraint set *Γ implies* a constraint *c* iff *c* is *true* whenever all constraints of *Γ* are *true*. For example, as acyclicity implies irreflexivity for any self-map *sm* (as any *sm*(x) = x corresponds to a cycle of length 0), the set {*sm* acyclic} implies the constraint *sm* irreflexive.
7. A constraint *c* is *redundant* in a constraint set *Γ* iff {*Γ* – *c*} implies c. For example, in the set {*sm* acyclic, *sm* irreflexive}, *sm* irreflexive is redundant, for any self-map *sm*.
8. A constraint set is *minimal* iff it does not contain any redundant constraint.

Obviously, any DBMS must accept only satisfiable and coherent set of constraints and should enforce only minimal ones. Moreover, single (i.e., not composite) totally defined reflexive self-maps should never be stored, as they would duplicate the unity mappings of the corresponding sets (and thus, the surrogate keys of the corresponding db tables).

In what follows, we consider any finite set *S* having at least 4 elements (which is a norm in dbs), any self-map *sm* over it, and any atomic mappings $f : A \to B$, $g : B \to C$, and $h : C \to D$; we also consider 3 additional system (i.e., automatically added and deleted only by *MatBase* and read-only for its users) mapping constraint types: *f* self-map, *f* canonical (Cartesian) projection, and *f* canonical injection. The following propositions and corollaries hold:

***Propositions***

0. (i) *sm* might be connected iff *S* has at most 3 elements
   (ii) *sm* idempotent ⇔ *sm* transitive
   (iii) *sm* anti-idempotent ⇔ *sm* intransitive
   (iv) *sm* idempotent and anti-idempotent ⇔ $sm^2(x) \in$NULLS, $\forall x \in S \Rightarrow$ *sm* null-idempotent.
   (v) *sm* might not be left-Euclidean as $x \neq y \neq z \neq x$ would be contradicted
   (vi) *sm* might not be right-Euclidean as functionality would be contradicted
   (vii) *sm* left-inEuclidean ⇔ *sm* irreflexive
   (viii) *sm* might not be right-inEuclidean as functionality would be contradicted
   (ix) *sm* Euclidean and inEuclidean ⇔ *sm* one-to-one
1. (i) (*f* one-to-one ⇒ ¬ (*f* non-prime)) ∧ (*f* non-prime ⇒ ¬ (*f* one-to-one))
   (ii) (*f* total ⇒ ¬ (*f* default)) ∧ (*f* default ⇒ ¬ (*f* total))
   (iii) any canonical Cartesian projection *f* is totally defined
   (iv) no canonical Cartesian projection *f* may be non-prime

(v) *f* self-map ∧ (*f* reflexive ∨ *f* irreflexive ∨ *f* symmetric ∨ *f* asymmetric ∨ *f* idempotent ∨ *f* equivalence ∨ *f* acyclic ∨ *f* representative system mapping) (i.e., only self-maps may have dyadic-type properties)
(vi) (*sm* reflexive ⇒ ¬ (*sm* irreflexive)) ∧ (*sm* irreflexive ⇒ ¬ (*sm* reflexive))
(vii) (*sm* symmetric ⇒ ¬ (*sm* asymmetric)) ∧ (*sm* asymmetric ⇒ ¬ (*sm* symmetric))
(viii) (*sm* total ⇒ ¬ (*sm* null-reflexive ∨ *sm* null-symmetric ∨ *sm* null-idempotent ∨ *sm* null-equivalence ∨ *sm* null-representative system mapping)) ∧ ((*sm* null-reflexive ∨ *sm* null-symmetric ∨ *sm* null-idempotent ∨ *sm* null-equivalence ∨ *sm* null-representative system mapping) ⇒ ¬ ( *sm* total))
(ix) no *sm* may be a canonical Cartesian projection (i.e., no relation may be recursively defined over itself)
(x) any canonical injection is totally defined, one-to-one, reflexive, and idempotent self-map.
(xi) no canonical injection may be onto (i.e., in (E)MDM, inclusion is irreflexive [1])

2. (i) *f* one-to-one ∧ *g* one-to-one ⇒ *g* ° *f* one-to-one
(ii) *g* ° *f* one-to-one ⇒ *f* one-to-one ∧ $g|_{Im(f)}$ one-to-one
(iii) *g* ° *f* one-to-one ∧ *f* onto ⇒ *g* one-to-one
(iv) *f* onto ∧ *g* onto ⇒ *g* ° *f* onto
(v) *g* ° *f* onto ⇒ *g* onto
(vi) *g* ° *f* onto ∧ *g* one-to-one ⇒ *f* onto
(vii) *h* ° *g* ° *f* onto ∧ *h* one-to-one ⇒ *g* onto
(viii) *f* ° *g* self-map ∧ *f* ° *g* reflexive ⇒ *f* onto
(ix) *f* ° *g* self-map ∧ *f* ° *g* idempotent ⇔ *g* ° *f* reflexive
3. (i) *sm* onto ∧ *sm* total ⇔ *sm* one-to-one ∧ *sm* total
(ii) $\boldsymbol{1}_S$ is total, one-to-one, reflexive, and idempotent
(iii) *sm* = $\boldsymbol{1}_S$ ⇔ *sm* equivalence
(iv) *sm* reflexive ⇔ *sm* = $\boldsymbol{1}_S$
(v) *sm* representative system mapping ∧ *sm* one-to-one ⇒ *sm* reflexive
(vi) *sm* one-to-one ∧ *sm* ≠ $\boldsymbol{1}_S$ ⇒ *sm* irreflexive ∧ ¬(*sm* idempotent)
4. *sm* symmetric ⇔ $sm^2 = \boldsymbol{1}_S$
5. *sm* acyclic ⇔ $sm^n(x) \neq x$, $n > 0$, natural
6. *sm* total ∧ *sm* idempotent ⇔ $sm^n(x) = sm(x)$, $n > 0$, natural
7. (i) *sm* asymmetric ⇒ *sm* irreflexive
(ii) *sm* anti-idempotent ⇔ *sm* irreflexive
(iii) *sm* acyclic ⇒ *sm* asymmetric ∧ ¬(*sm* idempotent)
8. *sm* irreflexive ∧ *sm* idempotent ⇒ *sm* asymmetric
9. *sm* symmetric ∧ *sm* idempotent ⇒ *sm* reflexive
10. *sm* asymmetric ∧ *sm* idempotent ⇒ *sm* acyclic
11. *sm* representative system mapping ⇒ *sm* idempotent
12. (i) *sm* null-reflexive ∧ *sm* total ⇔ *sm* reflexive
(ii) *sm* null-symmetric ∧ *sm* total ⇔ *sm* symmetric
(iii) *sm* null-idempotent ∧ *sm* total ⇔ *sm* idempotent
(iv) *sm* null-equivalence ∧ *sm* total ⇔ *sm* equivalence
(v) *sm* null-representative system mapping ∧ *sm* total ⇔ *sm* representative system mapping
13. (i) *sm* null-reflexive ⇒ *sm* one-to-one ∧ *sm* null-idempotent
(ii) *sm* null-representative system mapping ∧ *sm* one-to-one ⇒ *sm* null-reflexive
(iii) *sm* irreflexive ∧ *sm* null-idempotent ⇒ *sm* asymmetric
(iv) *sm* null-symmetric ∧ *sm* null-idempotent ⇒ *sm* null-reflexive
(v) *sm* asymmetric ∧ *sm* null-idempotent ⇒ *sm* acyclic

(vi) *sm* null-representative system mapping ⇒ *sm* null-idempotent

***Corollaries***

0. For self-maps:

(i) Connectivity, intransitivity, Euclideanity, and inEuclideanity are of no interest.
(ii) Transitivity is replaced by idempotency.
(iii) For the satisfiability, coherence, and minimality of such constraint sets, as null-P-type constraints behave exactly as the corresponding P-type ones, the P-type ones and the *Total* constraint are all that end-users need (i.e., for P ∈{reflexivity, symmetry, idempotency, equivalence, representative system mapping } the Graphic User Interface (GUI) of *MatBase* only needs checkboxes for every one of them and one for *Total*, i.e., there is no need for any checkbox of type null-P; e.g., if the reflexivity checkbox is checked and the totality one is not, then *MatBase* considers null-reflexivity, otherwise it considers reflexivity instead).

1. Consider any atomic mapping $f : D \to C$ and self-map $sm : S \to S$;
Any constraint set containing any of the following combinations of constraint types is incoherent:
(i) *f* total ∧ *f* default
(ii) (*f* one-to-one ∨ *f* bijective) ∧ *f* non-prime
(iii) *f* canonical projection ∧ (¬(*f* total) ∨ *f* non-prime)
(iv) ¬(*f* self-map) ∧ (*f* reflexive ∨ *f* irreflexive ∨ *f* symmetric ∨ *f* asymmetric ∨ *f* idempotent ∨ *f* equivalence ∨ *f* acyclic ∨ *f* representative system mapping)
(v) *sm* canonical injection ∧ (*sm* onto ∨ ¬(*sm* total) ∨ ¬(*sm* one-to-one) ∨ ¬(*sm* reflexive) ∨ ¬(*sm* idempotent) ∨ ¬(*sm* self-map))
(vi) *sm* reflexive ∧ *sm* irreflexive
(vii) *sm* symmetric ∧ *sm* asymmetric
(viii) *sm* total ∧ *sm* onto ∧ *sm* non-prime
(ix) *sm* self-map ∧ *sm* canonical projection

Any constraint set containing any of the following combinations of constraint types is not minimal:
(x) *f* one-to-one ∧ *f* onto ∧ *f* bijective (*f* bijective is redundant, i.e., *f* one-to-one ∧ *f* onto ⇒ *f* bijective)
(xi) (*f* one-to-one ∨ *f* onto) ∧ *f* bijective (*f* one-to-one or/and *f* onto are redundant, i.e., *f* one-to-one ∧ *f* onto ⇐ *f* bijective)
(xii) *sm* reflexive ∧ *sm* symmetric ∧ *sm* idempotent ∧ *sm* equivalence (*sm* equivalence is redundant, i.e., *sm* reflexive ∧ *sm* symmetric ∧ *sm* idempotent ⇒ *sm* equivalence)
(xiii) (*sm* reflexive ∨ *sm* symmetric ∨ *sm* idempotent) ∧ *sm* equivalence (*sm* reflexive or/and *sm* symmetric or/and *sm* idempotent are redundant, i.e., *sm* reflexive ∧ *sm* symmetric ∧ *sm* idempotent ⇐ *sm* equivalence)

2. Consider any mappings $f : A \to B$, $g : B \to C$, and self-map $sm = h \circ i : S \to S$;

(i) any constraint set containing *f* one-to-one ∧ *g* one-to-one ∧ $g \circ f$ non-prime is incoherent (i.e., *f* one-to-one ∧ *g* one-to-one ⇒ $g \circ f$ one-to-one).
(ii) any constraint set containing *f* one-to-one ∧ *g* one-to-one ∧ $g \circ f$ one-to-one is not minimal, as $g \circ f$ one-to-one is redundant (i.e., *f* one-to-one ∧ *g* one-to-one ⇒ $g \circ f$ one-to-one).
(iii) any constraint set containing $g \circ f$ one-to-one ∧ (*f* non-prime ∨ $g|_{Im(f)}$ non-prime) is incoherent.
(iv) any constraint set containing $g \circ f$ one-to-one ∧ (*f* one-to-one ∨ $g|_{Im(f)}$ one-to-one) is not minimal, as *f* one-to-one and $g|_{Im(f)}$ one-to-one are redundant (i.e., $g \circ f$ one-to-one ⇒ *f* one-to-one ∧ $g|_{Im(f)}$ one-to-one).
(v) any constraint set containing $g \circ f$ one-to-one ∧ *f* onto ∧ *g* non-prime is incoherent.
(vi) any constraint set containing $g \circ f$ one-to-one ∧ *f* onto ∧ *g* one-to-one one is not minimal, as *g* one-to-one is redundant (i.e., $g \circ f$ one-to-one ∧ *f* onto ⇒ *g* one-to-one).

(vii) any constraint set containing $g \circ f$ onto is not minimal, as $g \circ f$ onto is redundant (i.e., $f$ onto ∧ $g$ onto ⇒ $g \circ f$ onto).
(viii) any constraint set containing $g \circ f$ onto ∧ $g$ onto is not minimal, as $g$ onto is redundant (i.e., $g \circ f$ onto ⇒ $g$ onto).
(ix) any constraint set containing $g \circ f$ onto ∧ $g$ one-to-one ∧ $f$ onto is not minimal, as $f$ onto is redundant (i.e., $g \circ f$ onto ∧ $g$ one-to-one ⇒ $f$ onto).
(x) any constraint set containing *sm* reflexive ∧ $h$ onto is not minimal, as $h$ onto is redundant (i.e., *sm* reflexive ⇒ $h$ onto).
(xi) any constraint set containing *sm* reflexive ∧ $i \circ h$ idempotent is not minimal, as $i \circ h$ idempotent is redundant (i.e., *sm* reflexive ⇒ $i \circ h$ idempotent).
(xii) any constraint set containing *sm* idempotent ∧ $i \circ h$ reflexive is not minimal, as $i \circ h$ reflexive is redundant (i.e., *sm* idempotent ⇒ $i \circ h$ reflexive).

3. Consider any mappings $f : A \to B$, $g : B \to C$, and $h : C \to D$; then any constraint set containing $h \circ g \circ f$ onto ∧ $h$ one-to-one ∧ $g$ onto is not minimal, as $g$ onto is redundant (i.e., $h \circ g \circ f$ onto ∧ $h$ one-to-one ⇒ $g$ onto).

For all following corollaries, consider any set $S$ and self-map $sm : S \to S$;

4. (i) Any constraint set containing *sm* total ∧ *sm* one-to-one ∧ (*sm* onto ∨ *sm* bijective) is not minimal, as *sm* onto and *sm* bijective are redundant (i.e., *sm* total ∧ *sm* onto ⇔ *sm* total ∧ *sm* one-to-one).
(ii) Any constraint set containing *sm* total ∧ (*sm* onto ∨ *sm* bijective) must be replaced by *sm* total ∧ *sm* one-to-one (with *sm* onto and *sm* bijective being flagged as redundant).
5. (i) Except for $sm = \boldsymbol{1}_S$, no other totally defined self-map may be declared as equivalence (i.e., *sm* total ∧ $sm \neq \boldsymbol{1}_S$ ∧ *sm* equivalence is rejected, as there is no sense in duplicating $\boldsymbol{1}_S$).
(ii) Only composite or not totally defined self-maps may be declared as reflexive, as there is no sense in duplicating $\boldsymbol{1}_S$ (i.e., *sm* total ∧ *sm* single ∧ *sm* reflexive is rejected).
(iii) Only non-totally defined self-maps may be declared as one-to-one representative system mappings, as there is no sense in duplicating $\boldsymbol{1}_S$ (i.e., *sm* total ∧ *sm* one-to-one ∧ *sm* representative system mapping is rejected).
(iv) Only non-totally defined self-maps may be declared as symmetric and idempotent, as there is no sense in duplicating $\boldsymbol{1}_S$ (i.e., *sm* total ∧ *sm* symmetric ∧ *sm* idempotent is rejected).
(v) Any constraint set containing *sm* one-to-one ∧ *sm* idempotent is incoherent.
(vi) Any constraint set containing *sm* one-to-one ∧ *sm* irreflexive (with $sm \neq \boldsymbol{1}_S$) is not minimal, as *sm* irreflexive is redundant (i.e., *sm* one-to-one ∧ $sm \neq \boldsymbol{1}_S$ ⇒ *sm* irreflexive).
6. (i) Any constraint set containing *sm* asymmetric ∧ *sm* reflexive is incoherent.
(ii) Any constraint set containing *sm* asymmetric ∧ *sm* irreflexive is not minimal, as *sm* irreflexive is redundant (i.e., *sm* asymmetric ⇒ *sm* irreflexive).
(iii) Any constraint set containing *sm* acyclic ∧ (*sm* idempotent ∨ *sm* symmetric ∨ *sm* reflexive) is incoherent.
(iv) Any constraint set containing *sm* acyclic ∧ (*sm* asymmetric ∨ *sm* irreflexive) is not minimal, as *sm* irreflexive and *sm* asymmetric are redundant (i.e., *sm* acyclic ⇒ *sm* asymmetric).
7. (i) Any constraint set containing *sm* irreflexive ∧ *sm* idempotent ∧ *sm* symmetric is incoherent.
(ii) Any constraint set containing *sm* irreflexive ∧ *sm* idempotent ∧ *sm* asymmetric is not minimal, as *sm* asymmetric is redundant (i.e., *sm* irreflexive ∧ *sm* idempotent ⇒ *sm* asymmetric).
8. Any constraint set containing *sm* asymmetric ∧ *sm* idempotent ∧ *sm* acyclic is not minimal, as *sm* acyclic is redundant (i.e., *sm* asymmetric ∧ *sm* idempotent ⇒ *sm* acyclic).
9. Any constraint set containing *sm* representative system mapping ∧ *sm* idempotent is not minimal, as *sm* idempotent is redundant (i.e., *sm* representative system mapping ⇒ *sm* idempotent).
10. (i) Any constraint set containing ¬(*sm* total) ∧ *sm* reflexive ∧ *sm* non-prime is incoherent.
(ii) Any constraint set containing ¬(*sm* total) ∧ *sm* reflexive ∧ *sm* one-to-one is not minimal, as *sm* one-to-one is redundant (i.e., *sm* null-reflexive ⇒ *sm* one-to-one).

(iii) Any constraint set containing ¬(*sm* total) ∧ *sm* representative system mapping ∧ *sm* one-to-one ∧ *sm* irreflexive is incoherent.

(iv) Any constraint set containing ¬(*sm* total) ∧ *sm* representative system mapping ∧ *sm* one-to-one ∧ *sm* reflexive is not minimal, as *sm* reflexive is redundant (i.e., *sm* null-representative system mapping ∧ *sm* one-to-one ⇒ *sm* null-reflexive).

(v) Any constraint set containing ¬(*sm* total) ∧ *sm* symmetric ∧ *sm* idempotent ∧ *sm* irreflexive is incoherent.

(vi) Any constraint set containing ¬(*sm* total) ∧ *sm* symmetric ∧ *sm* idempotent ∧ *sm* reflexive is not minimal, as *sm* reflexive is redundant (i.e., *sm* null-symmetric ∧ *sm* null-idempotent ⇒ *sm* null-reflexive).

*MatBase* stores in its metacatalog these 10 above corollaries, as well as needed data on self-maps and atomic and composite mappings in the tables presented in the following subsections.

### *Table COROLLARIES*

Table *COROLLARIES* (see Figure 2) stores data about the corollaries on the coherence and minimality of constraint sets (a surrogate primary autogenerated key x, corollaries' types, names, bodies, book volume, subsection, and page number in which they appear in [9], etc.). *COROLLARIES* also stores data for all other 56 (E)MDM constraint types [1], not only for the 20 self-map and general function ones (see, e.g., [2]). Data from this table (which was manually entered) is used for providing users with context-sensitive questions, warnings, and error messages.

Please note in Figure 2 the selected lines from this table: they correspond to a 3rd corollary type (besides *Incoherence* and *Redundancy* [2]), namely *Rejection*, which directs *MatBase* to reject any such constraint combination, as the corresponding self-map would duplicate the unity mapping of the corresponding set.

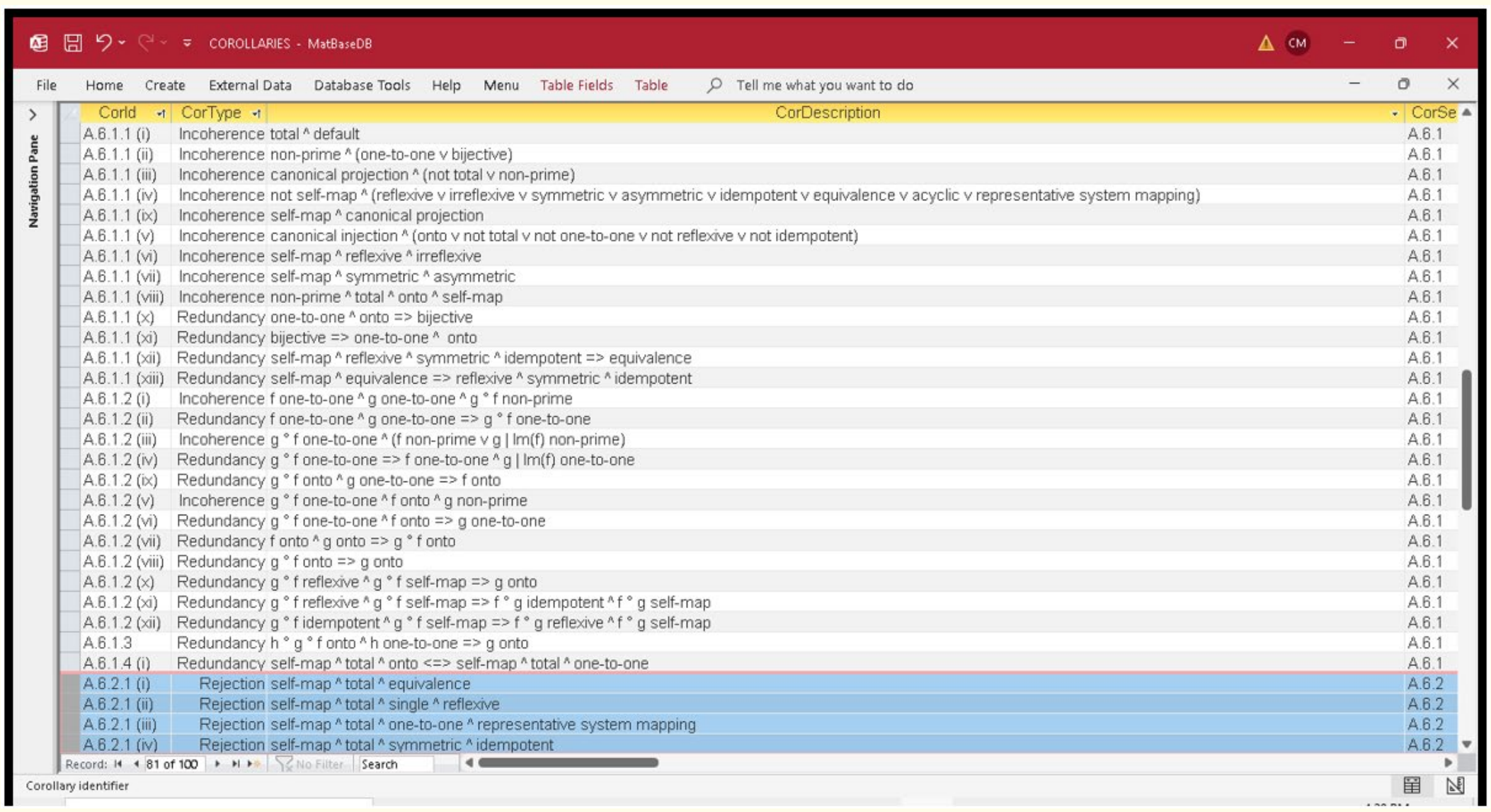

| CorId | CorType | CorDescription | CorSe |
|---|---|---|---|
| A.6.1.1 (i) | Incoherence | total ^ default | A.6.1 |
| A.6.1.1 (ii) | Incoherence | non-prime ^ (one-to-one v bijective) | A.6.1 |
| A.6.1.1 (iii) | Incoherence | canonical projection ^ (not total v non-prime) | A.6.1 |
| A.6.1.1 (iv) | Incoherence | not self-map ^ (reflexive v irreflexive v symmetric v asymmetric v idempotent v equivalence v acyclic v representative system mapping) | A.6.1 |
| A.6.1.1 (ix) | Incoherence | self-map ^ canonical projection | A.6.1 |
| A.6.1.1 (v) | Incoherence | canonical injection ^ (onto v not total v not one-to-one v not reflexive v not idempotent) | A.6.1 |
| A.6.1.1 (vi) | Incoherence | self-map ^ reflexive ^ irreflexive | A.6.1 |
| A.6.1.1 (vii) | Incoherence | self-map ^ symmetric ^ asymmetric | A.6.1 |
| A.6.1.1 (viii) | Incoherence | non-prime ^ total ^ onto ^ self-map | A.6.1 |
| A.6.1.1 (x) | Redundancy | one-to-one ^ onto => bijective | A.6.1 |
| A.6.1.1 (xi) | Redundancy | bijective => one-to-one ^ onto | A.6.1 |
| A.6.1.1 (xii) | Redundancy | self-map ^ reflexive ^ symmetric ^ idempotent => equivalence | A.6.1 |
| A.6.1.1 (xiii) | Redundancy | self-map ^ equivalence => reflexive ^ symmetric ^ idempotent | A.6.1 |
| A.6.1.2 (i) | Incoherence | f one-to-one ^ g one-to-one ^ g ° f non-prime | A.6.1 |
| A.6.1.2 (ii) | Redundancy | f one-to-one ^ g one-to-one => g ° f one-to-one | A.6.1 |
| A.6.1.2 (iii) | Incoherence | g ° f one-to-one ^ (f non-prime v g \| Im(f) non-prime) | A.6.1 |
| A.6.1.2 (iv) | Redundancy | g ° f one-to-one => f one-to-one ^ g \| Im(f) one-to-one | A.6.1 |
| A.6.1.2 (ix) | Redundancy | g ° f onto ^ g one-to-one => f onto | A.6.1 |
| A.6.1.2 (v) | Incoherence | g ° f one-to-one ^ f onto ^ g non-prime | A.6.1 |
| A.6.1.2 (vi) | Redundancy | g ° f one-to-one ^ f onto => g one-to-one | A.6.1 |
| A.6.1.2 (vii) | Redundancy | f onto ^ g onto => g ° f onto | A.6.1 |
| A.6.1.2 (viii) | Redundancy | g ° f onto => g onto | A.6.1 |
| A.6.1.2 (x) | Redundancy | g ° f reflexive ^ g ° f self-map => g onto | A.6.1 |
| A.6.1.2 (xi) | Redundancy | g ° f reflexive ^ g ° f self-map => f ° g idempotent ^ f ° g self-map | A.6.1 |
| A.6.1.2 (xii) | Redundancy | g ° f idempotent ^ g ° f self-map => f ° g reflexive ^ f ° g self-map | A.6.1 |
| A.6.1.3 | Redundancy | h ° g ° f onto ^ h one-to-one => g onto | A.6.1 |
| A.6.1.4 (i) | Redundancy | self-map ^ total ^ onto <=> self-map ^ total ^ one-to-one | A.6.1 |
| A.6.2.1 (i) | Rejection | self-map ^ total ^ equivalence | A.6.2 |
| A.6.2.1 (ii) | Rejection | self-map ^ total ^ single ^ reflexive | A.6.2 |
| A.6.2.1 (iii) | Rejection | self-map ^ total ^ one-to-one ^ representative system mapping | A.6.2 |
| A.6.2.1 (iv) | Rejection | self-map ^ total ^ symmetric ^ idempotent | A.6.2 |

***Figure 2:*** MS Access *MatBase COROLLARIES* table for storing corollaries on the coherence and minimality of constraint sets.

***Tables SMCCoherencies and SMCAdditionalRedund***

Table *SMCCoherencies* (see Figure 3) stores data about the coherency of the non-trivial self-map and general function constraint type combinations (out of the $2^{17} - 1$ = 131,071 possible ones). Abbreviations of the 18 columns of *SMCCoherencies* after the primary key *x* have the following meanings: *Ch* = Coherent?, *SM* = Self-map?, *CP* = Canonical projection?, *CI* = Canonical injection?, *RS* = Representative System mapping?, *A* = Acyclic?, *Q* = eQuivalence?, *I* = Idempotent?, *AS* = Asymmetric?, *S* = Symmetric? *IR* = Irreflexive?, *R* = Reflexive?, *B* = Bijective?, *OT* = Onto?, *UK* =Injective? (Unique Key?), *NP* = Non-Prime?, *DV* = Default Value?, *T* = Total?.

The unique combination numbers *x* are computed as the decimal equivalents of the corresponding binary ones, just like for all other tables storing constraint type combinations (where *SM* is multiplied by $2^{16}$ = 65536, CP by $2^{15}$ = 32768, …, and *T* by $2^0$ = 1, i.e., *x* = [*T*] + 2*[*DV*] + 4*[*NP*] + 8*[*UK*] + 16*[*OT*] + 32*[*B*] + 64*[*R*] + 128*[*IR*] + 256*[*S*] + 512*[*AS*] + 1024*[*I*] + 2048*[*Q*] + 4096*[*A*] + 8192*[*RS*] + 16384*[*CI*] + 32768*[*CP*] + 65536*[*SM*]).

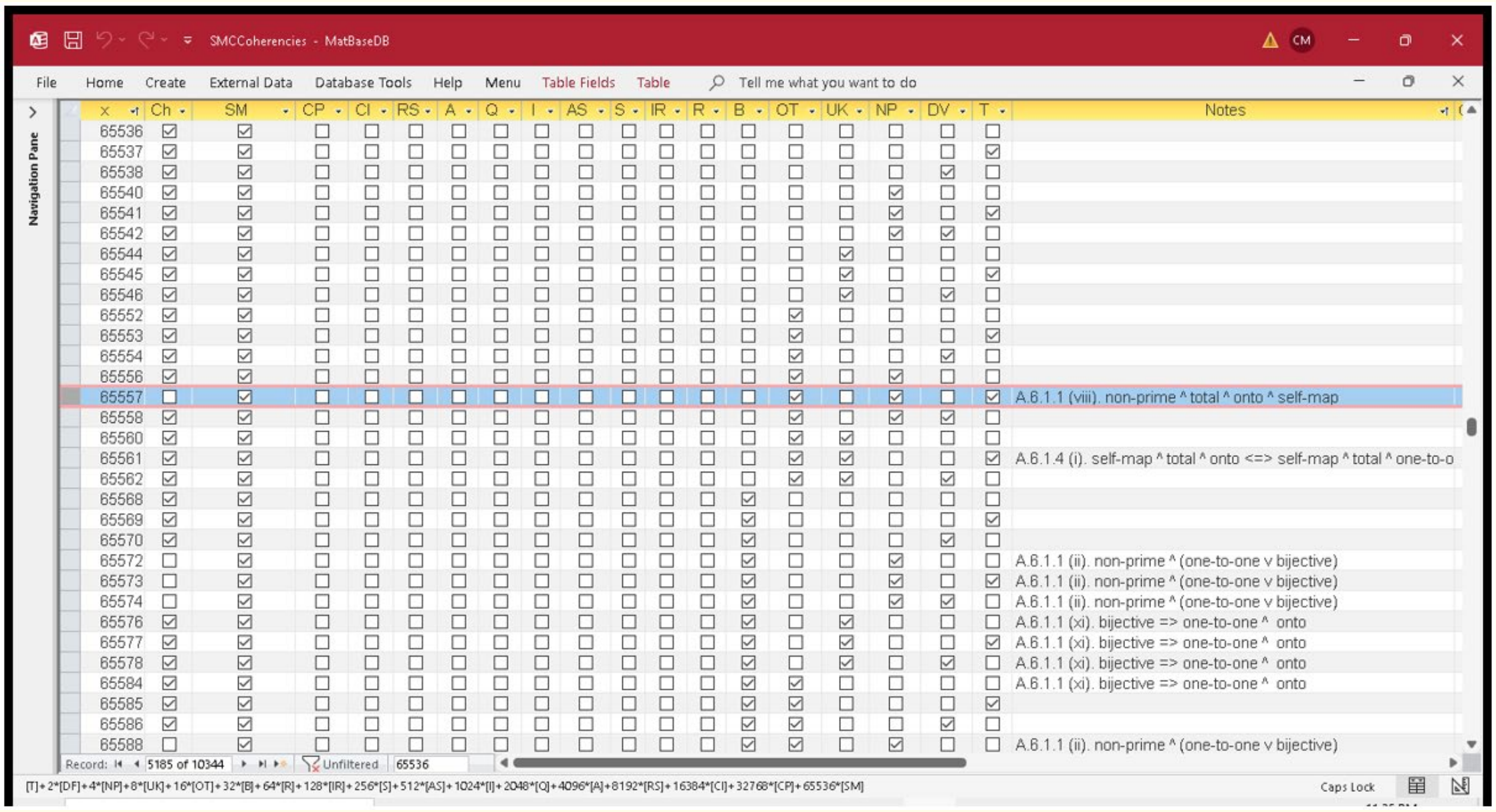

***Figure 3:*** MS Access *MatBase SMCCoherencies* table for storing non-trivial combinations of general mapping and self-map constraint types.

For example, combinations {Self-map, Onto} and {Self-map, Onto, Non-prime} have 65552 and 65556, respectively, as values for x (Self-map being multiplied by $2^{16}$, Onto being multiplied by $2^4$ and Non-prime by $2^2$) and are coherent, while the one for *x* = 65557, i.e., {Self-map, Onto, Non-prime, Total} is incoherent (as, according to Corollary 1(*viii*), any totally defined and onto self-map cannot be non-prime as well, because, according to Proposition 3(*i*), any such self-map is one-to-one as well, so, according to Proposition 1(*i*), it may not be non-prime).

Obviously, *Notes* is a foreign key referencing the primary key *x* of table *COROLLARIES*, from which its combo-box displays the corresponding values from the *CorId* and *CorDescription* columns for incoherent and not minimal combinations. The corresponding combo-box row source SQL statement is the following:

SELECT x, CorId & ". " & CorDescription AS [CorollaryID, Body] FROM COROLLARIES

WHERE CorSection like "A.6.*" ORDER BY CorId;

*SMCCoherencies* instance was automatically generated using SQL insert and update queries as follows: a query first inserted all non-trivial possible combinations (the trivial ones, i.e., those from Corollary 1(*i*) to (*ix*), are not stored); then, queries were run for each of the other 19 incoherence results, marking corresponding combinations as incoherent. For example, the query corresponding to Corollary 6(*iii*) is the following one (where 77 is the value of the primary key x for Corollary 6(*iii*) in table *COROLLARIES*):

UPDATE [SMCCoherencies] SET [Ch] = False, [Notes] = 77 WHERE [A] AND ([I] OR [S] OR [R]);

Finally, queries were run for all redundancy corollaries to update notes for the coherent but not minimal constraint set ones. For example, the query corresponding to Corollary 1(*xi*) is the following (where 99 is the value of the primary key x for Corollary 1(*xi*) in table *COROLLARIES*):

UPDATE [SMCCoherencies] SET [Notes] = 99 WHERE [Ch] AND [B] AND ([UK] OR [OT]);

Generally, more than one redundancy corollary may apply to a constraint set. For example, the set {Bijective, One-to-one, Onto} has both One-to-one and Onto redundant, according to corollary 1(xi). Consequently, there is also a table *SMCAdditionalRedund* in *MatBase's* metacatalog for storing the rest of redundancies for combinations having more than one; its structure is identical to the one of the table *SMCRedundancies* presented in the next subsection and its instance is also automatically populated with SQL INSERT statements. As this table is used only for automatically adding rows to the *SMCRedundancies* table and then for displaying accurate context-sensitive information and error messages, to keep things simple in this paper we are not providing more details on how it is used.

***Table SMCRedundancies***

Table *SMCRedundancies* (see Figure 4) stores data on the minimality of self-map and atomic map constraint type sets. Column *SMCCombination* is a foreign key referencing the primary key x of table *SMCCoherencies*; column *Notes* is just like the homonym one in table *SMCCoherencies*, except for the fact that it points to the subset of corollaries having type "Redundancy"; the corresponding combo-box row source SQL statement is the following:

SELECT x, CorId & ". " & CorDescription AS [CorollaryID, Body] FROM COROLLARIES

WHERE CorType = 1 AND CorSection Like "A.6.*" ORDER BY CorId;

Finally, the column *Redundancy* stores the redundant constraint types that make the corresponding constraint sets not minimal.

As an example, for any constraint set having *SMCCombination* = 65545, corresponding in table *SMCCoherencies* to the line having x = 65545, which encodes a constraint set of type {Self-map, One-to-one, Total}, there are two lines in table *SMCRedundancies* storing the fact that both bijectivity and ontoness must be added as redundant (see the selected lines from Figure 4).

The instance of *SMCRedundancies* was also automatically generated by running SQL queries for each redundancy corollary. For example, for corollary A.6.1.4 (i) (see the first row above the selected ones from Figure 2 and, e.g., row 65561 from Figure 3), the following two SQL statements were run for inserting the two selected lines from Figure 4:

INSERT INTO SMCRedundancies (SMCCombination, [Notes], Redundancy)

SELECT x, [Notes], "B" FROM SMCCoherencies WHERE [Ch] AND [SM] AND [UK] AND [T];

INSERT INTO SMCRedundancies (SMCCombination, [Notes], Redundancy)

SELECT x, [Notes], "OT" FROM SMCCoherencies WHERE [Ch] AND [SM] AND [UK] AND [T];

Please note that these SQL statements need to be recursively run up until no new redundancy is added to *SMCRedundancies*, just like it is the case for the dyadic relation constraints (see [2]).

***Tables DATABASES and CONSTRAINTSETS***

Figure 5 shows *MatBase's* metacatalog tables for storing data on the dbs and associated constraint sets it manages. Table *DATABASES* has a surrogate primary key *(#DB*), columns for storing the name (*DBName*), path where they reside on the server (*DBPath*), type (*DBType*), whether they are system (i.e., part of the metacatalog) or user ones (*System*), and a short description of what data each db stores (*DBSemantics*), as well as other needed details for their backup and restore.

Table *CONSTRAINTSETS* also has a surrogate primary key (*#C*), columns for storing the name (*ConstraintName*), db to which they belong (*Database*), type (*ConstraintType*), whether they are system (i.e., implicit ones, like *Self-map*, *Canonical projection*, *Canonical injection*) or user (explicit) ones (*System*), and a short description (*ConstraintSemantics*), as well as other equally important columns like *ImplyedBy* (for those that are implied by other constraints), *Set* and *Mapping* (for storing the object set or the mapping they constrain, if there is only one such set or mapping, respectively), the E-R Diagram cycle (see [9, 24]) to which it is associated, if any, etc. For example, Figure 5 shows the first columns of this table for the constraint set associated to a Geography db.

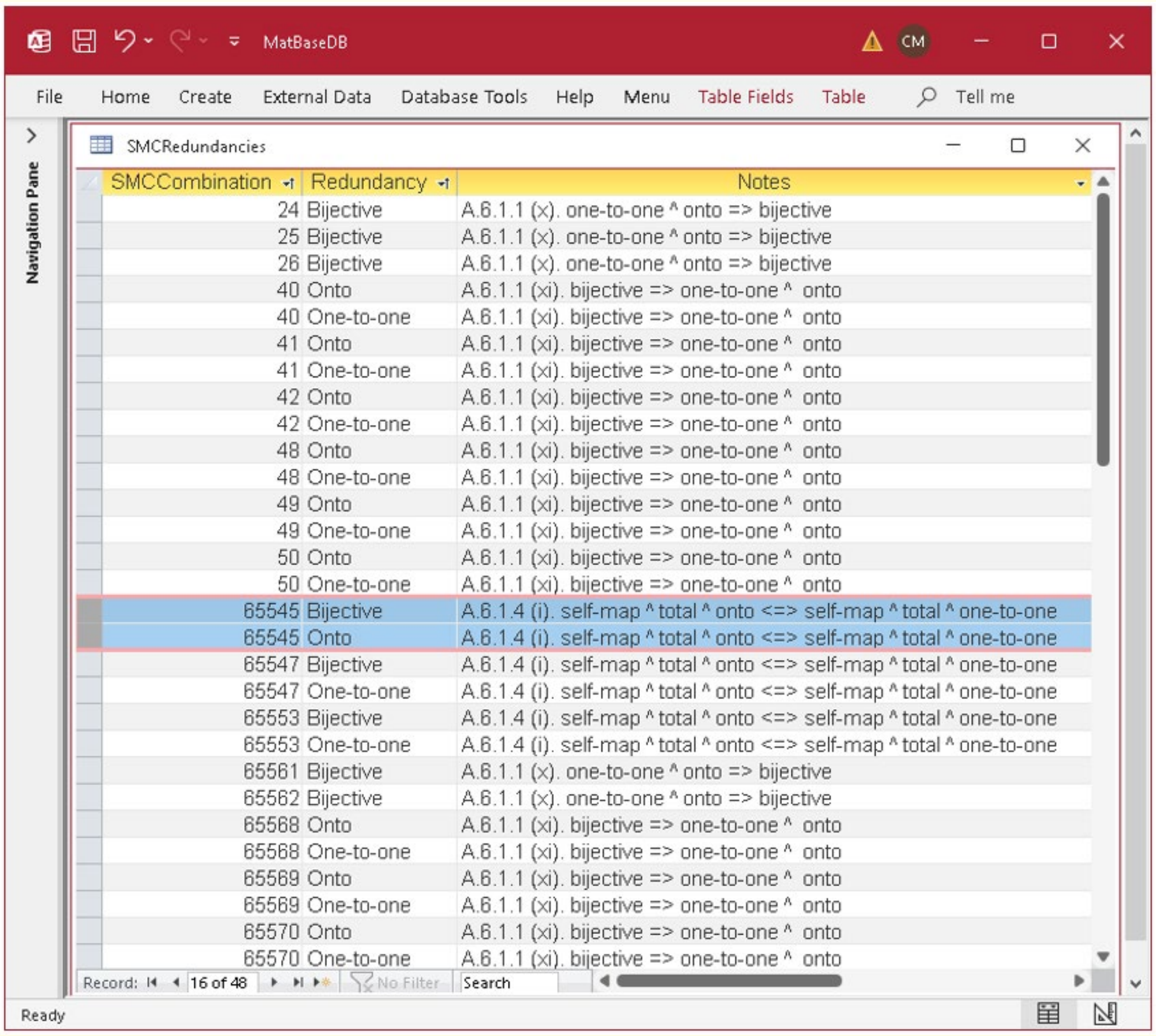

| SMCCombination | Redundancy | Notes |
|---|---|---|
| 24 | Bijective | A.6.1.1 (x). one-to-one ^ onto => bijective |
| 25 | Bijective | A.6.1.1 (x). one-to-one ^ onto => bijective |
| 26 | Bijective | A.6.1.1 (x). one-to-one ^ onto => bijective |
| 40 | Onto | A.6.1.1 (xi). bijective => one-to-one ^ onto |
| 40 | One-to-one | A.6.1.1 (xi). bijective => one-to-one ^ onto |
| 41 | Onto | A.6.1.1 (xi). bijective => one-to-one ^ onto |
| 41 | One-to-one | A.6.1.1 (xi). bijective => one-to-one ^ onto |
| 42 | Onto | A.6.1.1 (xi). bijective => one-to-one ^ onto |
| 42 | One-to-one | A.6.1.1 (xi). bijective => one-to-one ^ onto |
| 48 | Onto | A.6.1.1 (xi). bijective => one-to-one ^ onto |
| 48 | One-to-one | A.6.1.1 (xi). bijective => one-to-one ^ onto |
| 49 | Onto | A.6.1.1 (xi). bijective => one-to-one ^ onto |
| 49 | One-to-one | A.6.1.1 (xi). bijective => one-to-one ^ onto |
| 50 | Onto | A.6.1.1 (xi). bijective => one-to-one ^ onto |
| 50 | One-to-one | A.6.1.1 (xi). bijective => one-to-one ^ onto |
| 65545 | Bijective | A.6.1.4 (i). self-map ^ total ^ onto <=> self-map ^ total ^ one-to-one |
| 65545 | Onto | A.6.1.4 (i). self-map ^ total ^ onto <=> self-map ^ total ^ one-to-one |
| 65547 | Bijective | A.6.1.4 (i). self-map ^ total ^ onto <=> self-map ^ total ^ one-to-one |
| 65547 | One-to-one | A.6.1.4 (i). self-map ^ total ^ onto <=> self-map ^ total ^ one-to-one |
| 65553 | Bijective | A.6.1.4 (i). self-map ^ total ^ onto <=> self-map ^ total ^ one-to-one |
| 65553 | One-to-one | A.6.1.4 (i). self-map ^ total ^ onto <=> self-map ^ total ^ one-to-one |
| 65561 | Bijective | A.6.1.1 (x). one-to-one ^ onto => bijective |
| 65562 | Bijective | A.6.1.1 (x). one-to-one ^ onto => bijective |
| 65568 | Onto | A.6.1.1 (xi). bijective => one-to-one ^ onto |
| 65568 | One-to-one | A.6.1.1 (xi). bijective => one-to-one ^ onto |
| 65569 | Onto | A.6.1.1 (xi). bijective => one-to-one ^ onto |
| 65569 | One-to-one | A.6.1.1 (xi). bijective => one-to-one ^ onto |
| 65570 | Onto | A.6.1.1 (xi). bijective => one-to-one ^ onto |
| 65570 | One-to-one | A.6.1.1 (xi). bijective => one-to-one ^ onto |

***Figure 4:*** MS Access *MatBase SMCRedundancies* table for storing redundant constraint types of not minimal self-map constraint sets.

***Tables SETS, SetsCategories, and FUNCTIONS***

For any object set *S* that it manages, *MatBase* automatically generates its virtual unity mapping $\boldsymbol{1}_S$ and flags it as total, one-to-one, onto, bijective, reflexive, symmetric, idempotent, equivalence, and representative system mapping. Being automatically generated, the self-map $\boldsymbol{1}_S$ is a system object, which means that *MatBase* users may not either delete it or update its properties. Then, *MatBase* also generates its object-id called *x*. In the table storing *S's* instance, *x* is implemented as the surrogate primary key; if *S* is not a subset of another set, then *x's* values are autogenerated; if *S* is a subset of *T*, *x* is also the canonical injection associated to $S \subseteq T$, so it must take its values from *T's x* ones.

Whenever *S* is a relation (i.e., a db relationship-type object), *MatBase* users must specify its atomic canonical Cartesian projections, in any order but having distinct names. *MatBase* flags them as such and automatically declares them totally defined and adds to the db scheme their totally defined one-to-one Cartesian function product.

More attributes of the *MatBase's* metacatalog table *SETS* may be spotted in Figure 1: *SetName*, *Synonym* (of the set name), *SetType* (e.g., Entity, Relationship, Value, Calculated, System), *System*, *SetSemantics*, *card* (current set cardinal), *objectId* (*x*, generally, but db architects might rename it or choose another mapping instead), *SetCategory*, *SetCategOrdinal* (desired set position within its category), *StdUpdForm* (name of the standard software application Windows form managing set's data instance), *FactPredicate* (associated *Datalog¬* factual predicate), *minValue*, *maxValue* (minimum and maximum accepted values for its data instance), and *Static* (i.e., is the set a static one, like the rainbow colors set, or a dynamic one, to/from which users may add/delete elements?).

Figure 6 shows the MS Access *MatBase's* metacatalog table *SetsCategories* and, as an example, for the *Neighbors* category of the Geography db, some of its sets. Any set category belongs to a *Database* and a *SetCategory*, has a description (*SetsCategSemantics*) and a surrogate key value (*#SC*) and may be a *System* or user-defined one.

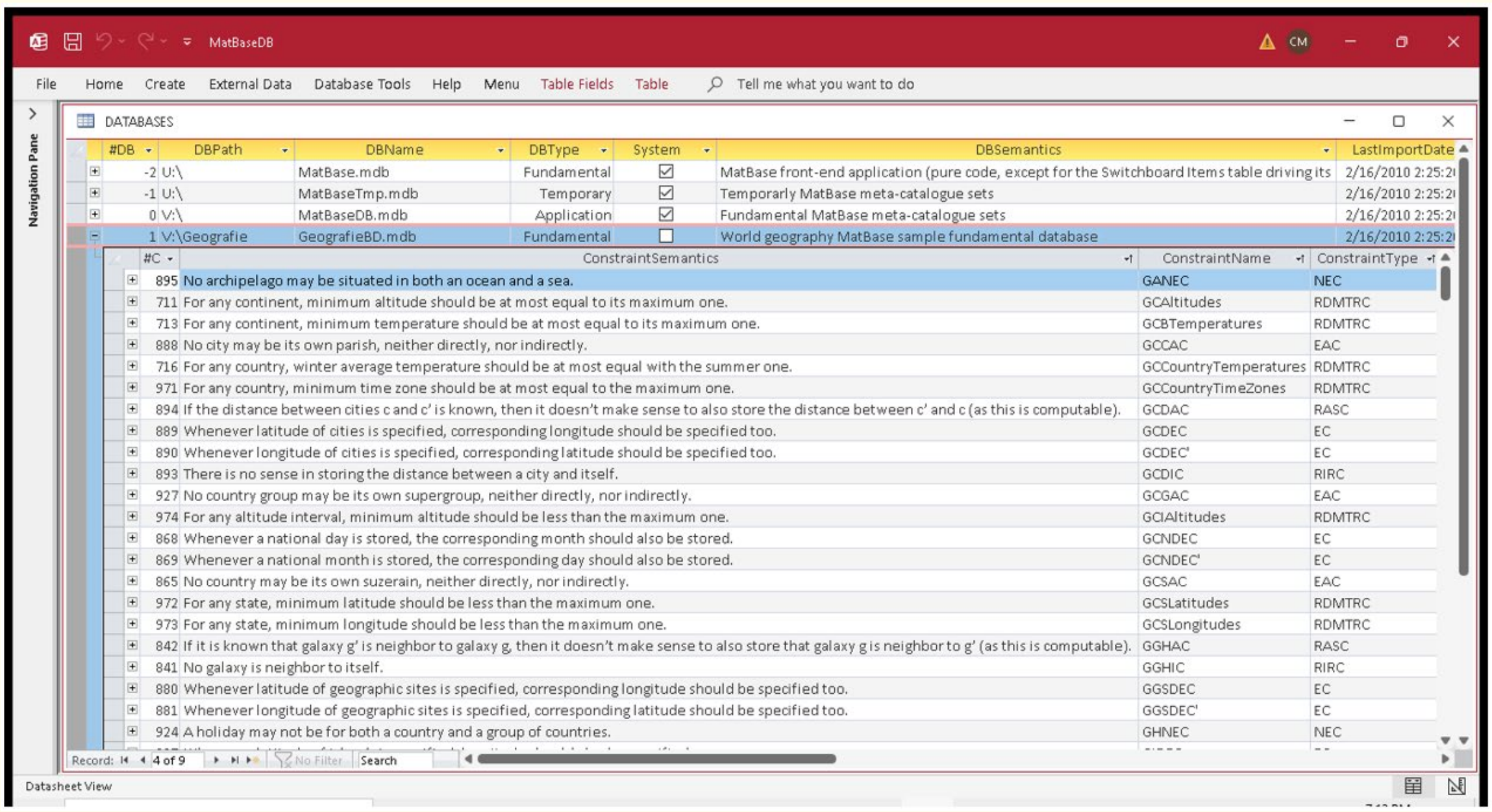

***Figure 5:*** MS Access *MatBase DATABASES* and *CONSTRAINTSETS* tables for storing metadata on managed databases and associated constraint sets.

Table *FUNCTIONS* stores data about the mappings managed by *MatBase*. For example, Figure 1 shows the ones defined on the set *STATES* from the *Geography* db: besides their implicit domain, please note their names, codomains, as well as some of their properties and constraints (*System, Total, Default value, Non-Prime, Injective, Surjective, Reflexive, Irreflexive, Symmetric, Asymmetric, Acyclic, RepresentativeSystemMapping*). By sliding the corresponding cursor bar to the right, *MatBase* users may inspect (and update the non-read-only ones) the rest of them (*Idempotent, Equivalence, CanonicalProjection, CanonicalInjection, Self-map, Arity, MinValue, MaxVal-*

*ue, Computed, Composed, System, FunctionSemantics*), as well as other ones that are irrelevant for this article.

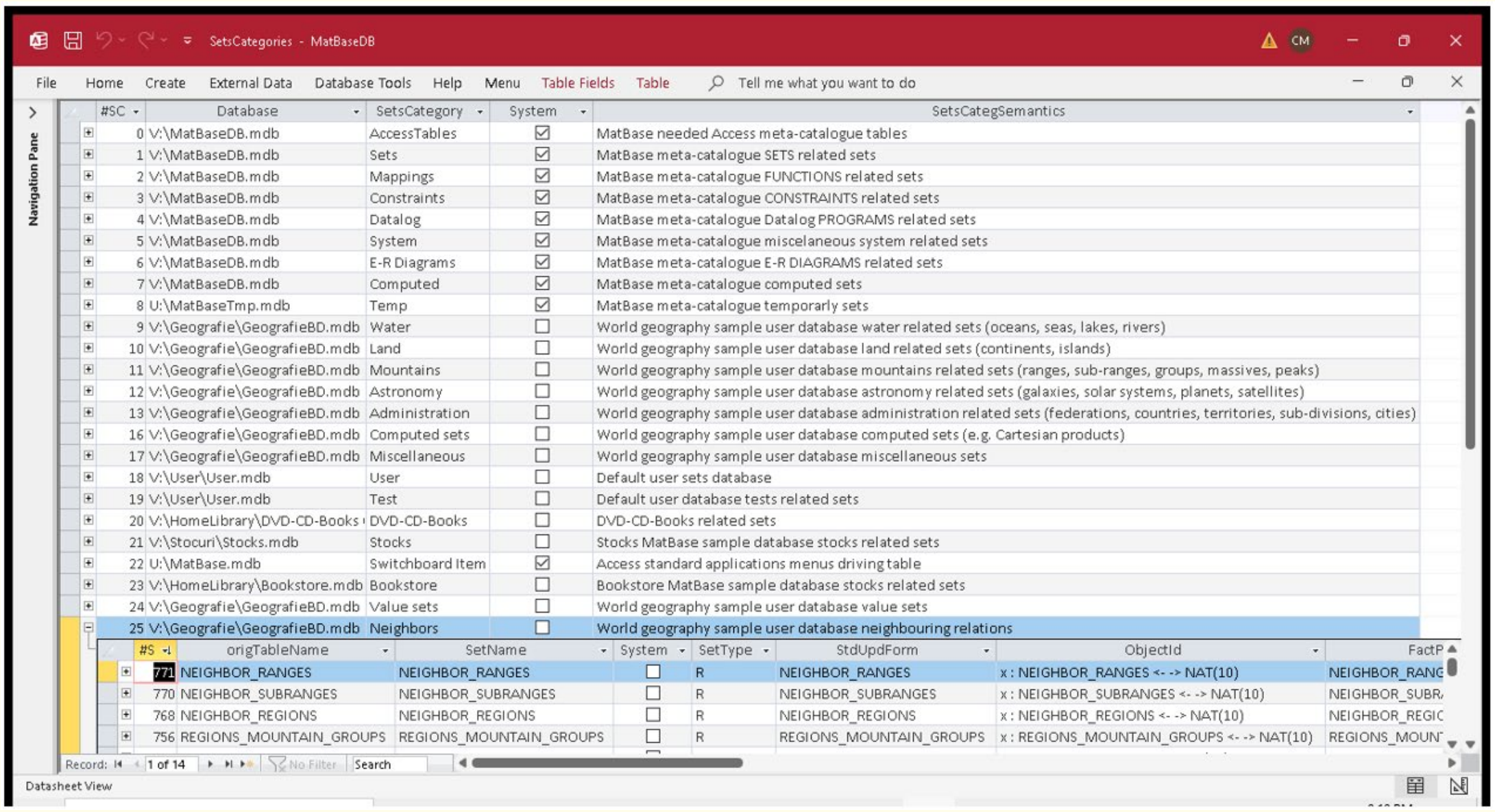

***Figure 6:*** MS Access *MatBase SetsCategories* table for storing metadata on managed set categories.

### *Tables *FUNCTIONS and COMP_FUNCT_COMP*

Composed mappings are a subset of the calculated ones, which are stored in the metacatalog table **FUNCTIONS* (in their turn, calculated mappings being a subset of all mappings stored in table *FUNCTIONS)*; consequently, the surrogate primary key of *COMP_FUNCT_COMP* is a foreign key referencing the corresponding one from table **FUNCTIONS* (which is a foreign key referencing the corresponding one from table *FUNCTIONS*). Please note from Figure 1 that, beside their name, domain, and codomain, calculated mappings (in this case the composed one *State* ° *StateCapital*) also have a mathematic *Formula* and a *SQL Formula*. The composed mappings are specified one by one, in the order (*Position*) of their composition (which is validated by *MatBase*: for any *g* ° *f*, *f 's* codomain must be equal to or a subset of *g*'s domain), and only accepting atomic mappings (*Member Function*).

### *The structural E-R diagram of this part of MatBase's metacatalog*

Figure 7 shows the structural E-R diagram of this part of *MatBase's* metacatalog, which makes crystal clear the functional relations between all the above 11 presented tables.

### *MatBase Algorithm SMCSCMEA*

When a user tries to add a new constraint *c* to a mapping *f* (defined over a set *S* and having an associated constraint set *C*) by clicking the corresponding checkbox shown in Figure 1 or one that becomes visible by sliding the corresponding cursor bar to the right, *MatBase* is first computing the *x* value of this new constraint set and is looking for it in table *SMCCoherencies*. If it doesn't find it, which only occurs when this set is trivially incoherent (e.g., *c* = Non-prime and *C* contains Injectivity), then it unchecks the corresponding box and displays the appropriate error message. If it finds it corresponding to an incoherent combination (e.g., {Self-map, Onto, Total} ∪ {Non-prime}), it rejects it as well, similarly. If the corresponding combination is coherent but has in table *SMCRedundancies* the type *Rejection* (i.e., the new constraint set corresponds to a unity mapping, see the selected lines from Figure 2), then it rejects it as well, similarly. Finally, in all other cases it checks whether the current *f*'s data instance satisfies *c* and if this is not the case it rejects *c* as well, similarly to the above cases.

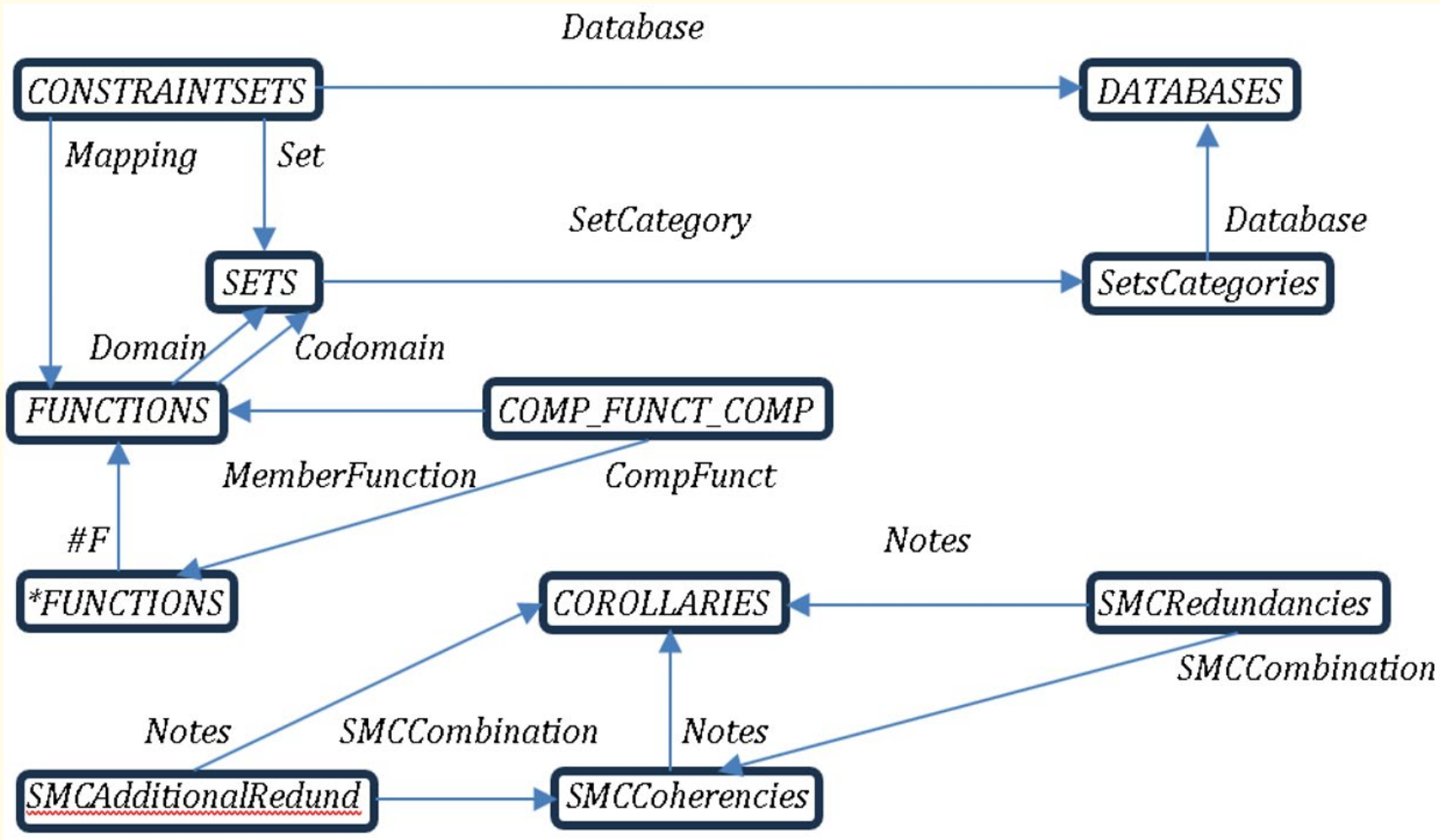

***Figure 7:*** The structural E-R diagram of the above *MatBase*'s metacatalog tables.

Whenever *c* is accepted both syntactically (i.e., from the coherence point of view) and semantically (i.e., from the data satisfiability one), *MatBase* adds to *S*'s form programming class (automatically generated immediately after table *S* has been added to the current db) calls to the corresponding *c* enforcement methods (which are publicly stored in its *Constraint* library) [10]. Moreover, if formerly not redundant constraints have become redundant, *MatBase* deletes from the *S*'s form programming class the code calling the corresponding public enforcement methods. Finally, it also automatically checks all newly redundant constraints, according to *SMCRedundancies* data for the newly *x* value from *SMCCoherencies* (e.g., if *C* = {Self-map, Total}, *c* = One-to-one, then *C*' = {Self-map, Total, One-to-one, Onto, Bijective}, with Onto and Bijective being both redundant).

For composed mappings things are slightly more complicated but similar. For example, if $f = g \circ h$, $C_f = \emptyset$, $C_g$ = {One-to-one}, $C_h$ ={Onto}, *c* = One-to-one, and *c* is satisfied by the current db instance (i.e., there are no duplicated *f*(*x*)), then $C_f$ ={One-to-one} but the one-to-oneness of *g* should not be enforced anymore, as it became redundant according to Corollary 2(*vi*).

Things are even more complicated when system constraints are automatically added. For example, let us consider $f : D \rightarrow E$, with $D \neq E$, $D \not\subset E$, $E \not\subset D$, and C = {Total, Onto}; if users successfully add the constraint $D \subseteq E$, *f* becomes a self-map, so, according to Corollary 4(*i*), *C* = {Self-map, Total, Onto, One-to-one, Bijective}, with ontoness and bijectivity redundant, which means that ontoness enforcing code should be removed and *f* one-to-one should be added (as it is simply enforceable through the underlying relational DBMS, while ontoness is not).

When a user tries to remove a constraint *c* by unchecking its corresponding non-read-only check-box, *MatBase* first computes the *x* value for the initial associated constraint set *C* and looks for *c* in *SMCRedundancies* table for *x*; if it finds it, then rejects the deletion attempt (as redundant constraints may not be deleted); otherwise, it removes from the *S*'s form programming class the calls to the constraint enforcement methods corresponding to *c*, then computes the corresponding new *x* value for *C*' and, finally, unchecks all formerly redundant constraints that are not implied anymore (e.g., if *C* = {Self-map, Acyclic, Asymmetric, Irreflexive} and *c* = Acyclic is deleted from it, then Asymmetric and Irreflexive are also deleted and *C*' = {Self-map}).

Similar to constraint additions, things are slightly more complicated but essentially the same for composed mappings. For example, if $f = g \circ h$, $C_f$ = {Onto}, $C_g$ = {One-to-one}, $C_h$ ={Onto}, and *c* = Onto, then $C_f = \emptyset$, $C_g$ = {One-to-one}, $C_h = \emptyset$, as, according to Corollary 2(*ix*), *h* Onto was redundant and now it is not anymore.

Again, things are even more complicated when system constraints are automatically removed. For example, let us consider $sm = g \circ f : S \to S$, $C$ = {Self-map, Total, One-to-one, Idempotent, Onto, Bijective}, with Onto and Bijective redundant, and $C_f$ = {One-to-one}; if users remove $g$ from table *COMP_FUNCT_COMP*, *sm* degenerates into an atomic single mapping, so that *MatBase* must also first remove $f$ from *COMP_FUNCT_COMP*, then *sm* from table **FUNCTIONS*, then *sm* from table *FUNCTIONS*, then drop *sm* together with its NOT NULL and UNIQUE constraints from the underlying relational db table *S*, then remove the code from *S*'s class for enforcing *sm*'s idempotency, then empty $C_f$ (as its one-to-oneness was redundant according to Corollary 2(*iv*)) and, if, in the corresponding db, there is a $rsm = f \circ g : S \to S$, remove from its constraint set the Reflexive constraint (as it was redundant, according to Corollary 2(*xii*)), and, finally, remove from S's class the code for enforcing *rsm*'s reflexivity.

Figure 8 presents the corresponding pseudocode algorithm used by *MatBase* to enforce self-map and atomic mapping constraints, while guaranteeing the satisfiability, coherency, and minimality of such constraint sets.

## Results and Discussion

***Proposition 14***

Algorithm *SMCSCMEA* from Figure 8 has the following properties:

(i) its complexity is a constant (i.e., $O(k)$)
(ii) it guarantees the satisfiability, coherence, and minimality of self-map, atomic, and composed mapping constraint sets
(iii) it is solid, complete, and optimal.

*Proof:*

(i) Trivially, it does not contain any loop, so it always ends in finite time after a (small) number of finite steps.

(ii) (*satisfiability*) Trivially, any void constraint set is satisfied by any data instance of any mapping and any non-void constraint set that is satisfied by a data instance remains satisfied after removing one of its constraints; as *SMCSCMEA* does not accept adding a new constraint to the constraint set of such a mapping if its instance does not satisfy it as well, it follows, obviously, that *SMCSCMEA* guarantees the satisfiability of such constraint sets.

(*coherence*) Trivially, any void constraint set is coherent, and any non-void coherent constraint set remains coherent after removing one of its constraints; as *SMCSCMEA* does not accept adding a new constraint to the constraint set of such a mapping if this would result in an incoherent set, it follows, obviously, that *SMCSCMEA* guarantees the coherence of such constraint sets as well.

(*minimality*) Trivially, any void constraint set is minimal; as *SMCSCMEA* is never enforcing redundant constraints but only signals them to the users for their info and is recomputing the subset of redundant constraints after accepting both adding and deleting a constraint, it follows, obviously, that *SMCSCMEA* guarantees the minimality of such constraint sets as well.

(iii) (*solidity*) Trivially, *SMCSCMEA* accepts to add to or delete from mapping constraint sets only the 23 mapping constraint types defined and characterized in the previous section.

(*completeness*) Trivially, *SMCSCMEA* accepts to add to or delete from mapping constraint sets all 23 types of mapping constraints defined and characterized in the previous section.

(*optimality*) Trivially, *SMCSCMEA* manages satisfiable, coherent, and minimal mapping constraint sets in the minimum possible number of steps, with the minimum possible accesses to the 11 tables presented in the previous section (and which are stored on external disks). *Q.E.D.*

The actual corresponding algorithms (written both in MS VBA and .NET C# with embedded SQL, respectively) are a little bit more complex, both to gain execution speed (by avoiding unnecessary disk reads), to prevent users from making unwanted mistakes, and to provide maximum possible accuracy for the context-sensitive messages it displays. For example, whenever the current mapping *f* has no constraints and the user adds one, it accepts it immediately if the current *f*'s instance satisfies it, as there may not be any corresponding either incoherency or redundancy. For example, whenever the user unchecks a constraint box, even if the corresponding deletion is possible *MatBase* displays a deletion confirmation message, does not proceed with the deletion if the request is not confirmed, and automatically unchecks the corresponding box. Moreover, if the request is confirmed and *c* is the only constraint of *C*, *MatBase* does not search for newly redundant constraints, as none may exist. Finally, automatically added or deleted system constraints are not dealt with by *SMCSCMEA*, as they are heavily dependent on other constraint types (e.g., set constraints) and composite mapping management.

**ALGORITHM *SMCSCMEA*** (Self-Map Constraint Sets Satisfiability, Coherence, and Minimality Enforcement)

***Input***: the current (E)MDM scheme constraint set *C* for an atomic mapping *f* over a set *S* and a constraint type *c* ∈ {total, default value, non-prime, one-to-one, onto, bijective, reflexive, irreflexive, symmetric, asymmetric, idempotent, equivalence, acyclic, representative system mapping} to be added to or removed from *C*.

***Output***: the updated coherent and minimal (E)MDM scheme *C '* for *f*.

***Strategy***:

*C '* = *C*; *ok* = *false*;
*if c* is to be removed from *C then* compute the *x* value for *C*;
  *if c*'s type is found redundant in a *SMCRedundancy* row having *SMCCombination* = *x then*
    display *c* & " cannot be removed as it is implied by other constraints, according to " & *Notes*(*x*);
  *else C '* = *C '* – { *c* }; *ok* = *true*;
    remove from the *S*'s class the calls to the constraint enforcement methods corresponding to *c*;
    compute the corresponding new *x* value for *C '*;
    remove from *C '* all formerly redundant constraints that are not implied anymore;
  *end if*;
*else* // *c* is to be added to *C*
  compute the corresponding *x* value for *C '* ∪ {*c*};
  *if* ¬*Ch*(*x*) *then* display *c* & " cannot be added, as, according to " & *Notes*(*x*) & ", the constraint set of " & *f* & " would become incoherent!";
  *elseif x* is missing from *SMCCoherencies then* display *c* & " cannot be added, as the constraint set of " & *f* & " would become incoherent!";
  *elseif SMCRedundancy* has a row having *SMCCombination* = *x* and *CorType* = "Rejection" *then*
    display *c* & " cannot be added, as, according to " & *Notes*(*x*) & ", " & *f* & " would become a unity mapping!";
  *else*
    *if c* is not satisfied by the current db instance *then* display *c* & " cannot be added to the constraint set of " & *f* & ", as its current instance does not satisfy it!"
    *else C '* = *C '* ∪ { *c* }; *ok* = *true*;
      add to the *S*'s class the calls to the corresponding constraint type enforcement methods;
      mark as true all redundant constraints that are implied by *C '* according to *SMCRedundancies*;
      remove from the *S*'s class the enforcement code for any constraint that just became redundant;
    *end if*;
  *end if*;
*end if*;
*if ok and f* is a compound mapping or a member of a compound one *cm then*
  correspondingly update constraint sets of *f*'s member functions or the one of *cm*, respectively;
*end if*;
***End*** ALGORITHM ***SMCSCMEA***;

***Figure 8:*** *MatBase* pseudocode Algorithm *SMCSCMEA*.

## Conclusion

We provided concise but accurate mathematical definitions for mappings, self-maps, their properties viewed as constraint types from the db perspective, as well as for the satisfiability, coherence, and minimality of such constraint sets.

We presented and discussed the pseudocode algorithm used by *MatBase* (our intelligent DBMS prototype based on both the relational, E-R, and our (Elementary) Mathematic data models) for enforcing atomic, composite, and self-map mapping constraint types, by guaranteeing the satisfiability, coherence, and minimality of such constraint sets. We also included description of the tables from *MatBase's* metacatalog needed for managing the corresponding metadata.

We proved that this algorithm actually guarantees both satisfiability, coherence, and minimality, while being fast, solid, complete, and optimal.

Obviously, the ultimate goal of the design and development of dbs and db software applications is to provide customers, first of all, with the tools that are not only user-friendly, but, above all, guaranteeing the highest possible data quality for their dbs and information extracted from them. If these tools do not guarantee the satisfiability and coherence of the associated constraint sets (be them enforced at the db or/and at the db software application levels), then junk data might (accidentally or purposely, it does not matter) be stored in their dbs, which leads to junk information extracted from them. Moreover, if these constraint sets are not minimal (which, it is true, does not impact data quality), then the corresponding db software applications run unnecessarily slower, to the dissatisfaction of their customers.

This paper also proves once more the formidable power of using mathematics (in particular, the naïve theory of sets, relations, and functions coupled with the first-order predicate calculus with equality) in dbs and db software applications design and development.

## Conflict of interest

The author declares that the research was conducted in the absence of any commercial or financial relationships that could be construed as a potential conflict of interest.

## Acknowledgements

This research was not sponsored by anybody and nobody other than its author contributed to it.